\documentclass[final,1p,times]{elsarticle} 
\usepackage{graphicx}
\usepackage{amssymb} 
\usepackage{amsthm} 
\usepackage{lineno}

\journal{Nuclear Physics A} 

\begin{document}

\begin{frontmatter} 

% Your Title - please insert
\title{Recent Theoretical Developments in Strongly Coupled QCD}

%% Single author (and collaboration) - please insert
\author{Ho-Ung Yee}
%\fntext[col1] {A list of members of the EMPIRE Collaboration and acknowledgements can be found at the end of this issue.}
\address{Department of Physics, University of Illinois, Chicago, Illinois 60607}\address{ and }\address{ RIKEN-BNL Research Center, Brookhaven National Laboratory, Upton, New York 11973-5000}

%% Multiple authors
%\author[auth2]{Marcus Junius Brutus}
%\address[auth1]{Somewhere, Rome}
%\address[auth2]{Somewhere else, Rome}

\begin{abstract} 
    Heavy-ion collisions involve strongly coupled dynamics of QCD in the entire history of time evolution. We
 review recent theoretical efforts to meet this challenge, focusing on the two approaches that the speaker has contributed to: 1) Holography or AdS/CFT correspondence, and 2)
 Symmetry protected phenomena such as those originating from triangle anomaly.
 The presentation is oriented to non-experts on these fields, and hence relies on intuitive pictures of the methods and the results, without going into specific details.

\end{abstract} 

\end{frontmatter} % do not change

%% linenumbers are useful for reviewing process
%\linenumbers

\section{Introduction}

Heavy-ion collisions involve many different aspects of strongly coupled dynamics of QCD in the entire evolution history of the created fireball.
To understand the initial multiplicity generation from the two colliding heavy nuclei, one needs to study the high energy scattering problem
in a wide range of momentum transfer $t$, and the physics in the small $t\ll \Lambda_{QCD}^2$ region  is non-perturbative and strongly coupled.
Hydrodynamic simulations successfully explaining the observed elliptic flow suggest that the thermalization of the resulting gluonic configuration to form the Quark-Gluon Plasma (QGP) happens faster than the time scale naively expected from pQCD.
These simulations also indicate a low value of shear viscosity which can not easily be explained in the weak coupling framework.
Given the strongly coupled nature of the QGP created, one has to deal with how the jets promptly created in the initial collision go through the strongly coupled QGP medium loosing 
a fraction or the entire of their energy to the medium.
As the QGP cools down, the phase transition to the hadronic phase giving us the final hadronic spectrum is well inside a realm of non-perturbative, strongly coupled QCD.

A well-accepted (and well-justified) strategy to meet this theoretical challenge is to use properly developed effective theories for each stage of the time history of the fireball evolution.
Another strategy available at the moment is to use an approximate model to QCD which shares all essential features of QCD, but nevertheless is solvable either analytically or numerically.
One example that will be reviewed in this talk is the AdS/CFT correspondence\cite{Maldacena:1997re}, based on the holographic duality between a strongly coupled, large $N_c$ limit of gauge theory (such as large $N_c$ QCD in strongly coupled regime) and a weakly coupled classical gravity in a 5-dimensional space (typically $AdS_5$ space).
The duality has been tested in many convincing ways only for certain supersymmetric gauge theories, but the underlying idea of holography is a much general concept beyond
the existence of supersymmetry, and there are several phenomenologically successful holographic models for QCD\cite{Sakai:2004cn,Erlich:2005qh}.

A somewhat orthogonal way to overcome the difficulty in dealing with strongly coupled dynamics of QCD is to look for the phenomena that are protected by
the symmetries of QCD. Low energy theorems based on the approximate chiral symmetry are good examples.
Although the conservation of approximate chiral symmetry doesn't seem very useful in the almost neutral QGP in the heavy-ion collisions,
the triangle anomaly of axial symmetry does introduce several new transport phenomena in the QGP phase, that lead to interesting experimentally-observable consequences, especially in the presence of the magnetic field
created in off-central collisions whose strength can be as large as $eB\sim m_\pi^2$. In this talk, we will review two such transport phenomena, the Chiral Magnetic Effect (CME)\cite{Kharzeev:2007jp,Fukushima:2008xe,Son:2004tq} and 
the Chiral Magnetic Wave (CMW)\cite{Kharzeev:2010gd,Newman:2005hd}, and their possible experimental signatures in heavy-ion collisions.

\section{Holography}

In this section, we try to give a short review on the recent developments of the application of holography to heavy-ion collisions,
selecting three topics 1) high energy scattering, 2) initial thermalization, and 3) energy loss of an energetic jet. 
The purpose of this review is to give the non-experts in the audience a large scale view on what have been achieved and what are the loose ends at the moment, without
going into much specific details. The presentation will be largely intuitive, focusing on the physics picture of the methods and the results, and hopefully this may give
us a good prospect on possible future development of the subjects. 

Before discussing the specific topics chosen, let us summarize a few important ideas (for non-experts in AdS/CFT) behind the holographic approach.
The holography is a conjectured duality between a gauge theory in the usual  4D Minkowski  space and a gravity theory in a 5D space having an extra "holographic dimension".
The extra dimension should be thought of as a geometric realization of energy scale of the gauge theory, and it is more like an interval (from UV to a finite IR) rather than a freely accessible extra dimension. Holographic 5D theory may be thought of as a foliation of 4D theories at different renormalization scales, where each energy scale corresponds to a point in the holographic direction. The entirety of the 5D theory is dual to the 4D gauge theory, and it is a bit misleading to think of the 4D gauge theory living at the UV boundary of the 5D space.
It is often useful in many applications to think of the holographic direction as representing the virtuality $Q=\sqrt{k^2-\omega^2}$ when $k>\omega$.
The gravity degrees of freedom in the 5D theory correspond to color neutral gauge invariant dynamics of the gauge theory, and it is generally hard to see colored dynamics directly. Colored charges are described as end points of strings moving in the 5D, which are dual to gauge theory flux tubes.
The gravity coupling in 5D is $G_N={1\over N_c^2}$, so a large $N_c$ limit of the gauge theory is a useful weak coupling limit in the 5D. Hence, the duality is useful in the large $N_c$ limit of the gauge theory, and gravity degrees of freedom can be thought of as large $N_c$ master fields.

A finite temperature, deconfined phase of the gauge theory is represented by a black-brane solution of the 5D gravity theory\cite{Witten:1998qj}, whose horizon is located at some point in the holographic direction and spans the whole spatial $R^3$. The Hawking temperature of the black brane is naturally mapped to the temperature of the gauge theory. Hawking radiations from the horizon are reflected back by the UV boundary to the horizon again, forming a dynamically stable equilibrium situation. Thermal fluctuations of the gauge theory are described by the fluctuations induced by these Hawking waves in the 5D. The area of the black brane horizon gives us the entropy of the gauge theory via the Hawking's area law; $S={A\over 4 G_N}\sim N_c^2$.
Cooling of the plasma maps to a receding black brane horizon toward the more IR region in the holographic direction.
Let us now discuss our specific topics in detail.

\subsection{High $s$, small $t$ scattering}

To appreciate the non-perturbative nature of the high $s$, small $t$ scattering, imagine two protons separated by a distance $b\gg \Lambda_{QCD}^{-1}$.
When they are static, we wouldn't expect any interaction between them due to confinement. Now, suppose  they are moving with a large relative rapidity $\chi\sim \log s$ with
an impact parameter $b$ so that they are always separated by at least a distance $b$. Naively, one might still expect their interaction to be negligible due to confinement since they are
separated at least by $b\gg \Lambda_{QCD}^{-1}$. The resulting total cross section would not be greater than $\Lambda_{QCD}^{-2}$ set by the confinement length scale.
This expectation is however challenged by the experiments showing a power-like growth of the total cross section $\sigma_T\sim s^{0.08}$\cite{Donnachie:1992ny}, indicating that  a large $s$ scattering involves
an interaction whose length scale grows with $s$ beyond the QCD scale, and hence is in the regime of non-perturbative QCD. What is responsible for this interaction has been the subject of many years of studies. The mediators of this interaction phenomenologically introduced are called Pomerons (and Reggeons for flavor changing cross sections).

For a large $(-t)\gg \Lambda_{QCD}^2$ (equivalently $b\ll \Lambda_{QCD}^{-1}$), one can study high $s$ scattering in pQCD and get useful intuition on the problem; this was done by BFKL\cite{BFKL}.
BFKL result gives a power-like growing of the total cross section, but more importantly it provides us with a physics picture why this happens.
A proton moving with a large $\chi$ emits virtual gluons carrying a fraction of the proton's momentum. These gluons successively emit daughter gluons carrying fractions of the momenta
of the mother gluons, and these steps can continue until the momenta of the gluons become comparable to an IR cutoff. The original proton is effectively
accompanied by a cloud of gluons that are produced by these successive branchings. These gluons can be usefully described as color dipoles\cite{Mueller:1993rr}.
Their transverse spatial distribution can be described by a diffusion type equation whose effective time
is naturally the rapidity $\chi$ that governs the number of branching steps: the larger $\chi$, the more the number of steps, and hence the bigger transverse radius of the gluon cloud leading to a growing total cross section.

There are indications that a similar type of diffusion may also exist in non-perturbative regime $(-t)\ll \Lambda_{QCD}^2$.
A Fourier transform of the Regge behavior of the $t$ dependence of the scattering amplitude $A\sim s^{\alpha_0+\alpha' t}$ into the impact parameter space $b$ gives us
$A\sim e^{-{b^2\over 2\alpha' \chi}}$, which is a solution of the 2D diffusion equation with the diffusion constant $D={\alpha'\over 2}$ and the effective time $\chi$.
This indicates that there are new degrees of freedom that continue sharing the momentum fractions of the proton below $\Lambda_{QCD}$.
Presumably, the soft pQCD gluons form color neutral bound states below $\Lambda_{QCD}$ and these object continue to diffuse in the transverse space. What has changed from pQCD
is simply the diffusion constant reflecting the branching strength of these objects. 
The old string theory as an effective theory of QCD flux tubes in confinement regime can reproduce these aspects, which was highlighted by the Veneziano amplitude.
The glue ball states obtained by quantizing the flux tubes are the objects that diffuse carrying fractions of the proton momentum. Note that these states should be massless off-shell states
located at $t=0$ point of the Regge trajectory, in order to be effective in the large length scales.
However, the string theory in the flat 4D space-time is found to be inconsistent, and has been abandoned as a theory of QCD.

In the framework of AdS/CFT, the string theory picture revives as a holographic dual theory of QCD, living not in 4D but in the holographic 5D space-time.
Stringy Pomeron in AdS/CFT provides us with a consistent picture of high $s$ scattering in low $(-t)$ regime\cite{Rho:1999jm,Janik:1999zk,Brower:2006ea,Basar:2012jb}.
The degrees of freedom in the 5D theory obtained by quantizing the strings represent color neutral degrees of freedom in confinement regime,
and the massless off-shell state at $t=0$ point of the 5D Regge trajectory indeed diffuses and governs the total cross section.
Because of the existence of the holographic direction, the diffusion happens not only in the transverse space, but also along the holographic direction, making the diffusion equation effectively 3-dimensional. Since holographic direction maps to the virtuality $Q$, the holographic dipole distribution can be thought of as encoding the dependence of the distribution on the virtuality. In a conformal regime, the resulting 3D distribution\cite{Stoffers:2012zw} is surprisingly similar to the BFKL-Mueller distribution of BFKL dipoles\cite{Mueller:1994gb}.

The behavior $A\sim e^{-{b^2\over 2\alpha' \chi}}$ can be understood as a string instanton amplitude. The explicit instanton solution indicates a strong analogy to the instantons in the Schwinger mechanism of pair creation in the presence of an electric field, which in our case should be induced by the rapidity. The fact that Regge amplitude can be described by an instanton means that the high $s$ scattering can be viewed as a tunneling of color charges through a wall of confining medium in the vacuum. From the instanton solution, one finds a non-zero Unrue temperature due to acceleration by the electric field, which can create a micro fireball\cite{Basar:2012jb}. This may explain the observed thermal nature of multiplicity from the high energy pp collisions.

The power-growing cross section eventually violates the unitarity bound $\sigma_T\le (\log s)^2$. One can achieve a reconciliation with the unitarity bound by eikonalizing the
single Pomeron amplitude: $A_1\to A_{eikonal}\sim N_c^2(1-e^{-{1\over N_c^2}A_1})$, where $A_1$ is the single Pomeron amplitude. 
As seen from the $N_c$ counting, this procedure involves adding up all higher string loop diagrams in the 5D space which are suppressed by powers of $1\over N_c^2$.
The regime where these correction become important is analogous to the saturation regime in pQCD where non-linear effects are no longer negligible. We currently don't know how to handle these multi-Pomerons in AdS/CFT to satisfy the unitarity bound, which is an important open problem. A dimensional reduction, something similar to what Refs.\cite{Lipatov:1991nf,Verlinde:1993te} found in pQCD, might be useful to solve this problem in AdS/CFT. The conflict to the unitarity bound at leading $N_c$ indicates that the high energy limit does not commute with the large $N_c$ limit.
We will arrive at the same conclusion when we discuss jet quenching below.

\subsection{Thermalization}

A fast thermalization time, $\tau\le 0.5$ fm, inferred by hydrodynamic fits to the measured elliptic flows requires its explanation in the strongly coupled dynamics of QCD.
The problem is to construct the 5D holographic state which is out-of-equilibrium, mimicking the initial condition right after the heavy-ion collision, and to study its time evolution toward 
a late-time black brane solution representing thermally equilibrated QGP. There are currently three types of 5D states considered for this purpose: 1) Two light-like gravitational shock waves in $AdS_5$ colliding head-on to each other\cite{Janik:2005zt,Albacete:2008vs,Chesler:2010bi,Gubser:2008pc,Wu:2011yd}, 2) Spatially homogeneous mass shell falling along the holographic direction from the UV region toward the IR region\cite{Lin:2006rf,Balasubramanian:2010ce}, and 3) Boost invariant solutions of the 5D Einstein equations with various initial conditions\cite{Beuf:2009cx,Heller:2011ju}. 

In the type 1) solution, the heavy-ions are modeled by energetic shock waves in 5D that collide with each other to form a black-hole.  One can numerically study the thermalization time and the multiplicity generation. The latter information is read off from the solution via the area of the resulting black-hole, since the entropy is given by the area of the black-hole; $S={A\over 4 G_N}$.
The type 2) solution has an advantage of having a relatively compact analytic expression for the solution, although what the falling in the holographic direction means in the gauge theory side and what is the proper initial condition to start with are not completely clear. As the holographic direction maps to the virtuality, the falling may represent the relaxation of initial high virtuality of the system constituents to a more on-shell thermal distribution. In this interpretation, a natural initial condition would be given by the saturation scale $Q_s$. In any case, one can study thermalization of various correlation functions such as energy-momentum two-point function and the entanglement entropy, to estimate the thermalization time.
The type 3) approach is to solve the 5D Einstein equation with various boost-invariant initial conditions, hoping to extract some universal properties in how the system thermalizes.
What has been found is that the system starts to be described well by the viscous hydrodynamics much before the isotropization is achieved. This effective thermalization time
is comparable to the time scale set by the final equilibrium temperature.

The common outcome from the above three approaches is that the thermalization indeed happens fast, meaning that the thermalization time is an order one number in unit of the final temperature, without any parametrical separation as it happens in pQCD.

\subsection{Jet quenching}

In the 5D holographic space, a strongly coupled QGP is described by a black brane solution. Since a colored charge is represented by an end point
of a 5D string, an energetic colored jet in QCD penetrating through the QGP would correspond to an end point of a 5D string moving through the background of the black brane.
Since the original jet is highly virtual, the initial position of the string end point in the holographic direction should be located in the UV regime, and falling toward the IR regime
as time goes on corresponds to the relaxation of virtuality via fragmentation. It is however hard to see the microscopic pattern of fragmentation in AdS/CFT directly.
The quest is to find the energy scaling of the penetration length of this holographic jet moving through a strongly coupled QGP.
As it turns out, the fact that the jet is color charged (and hence accompanies a 5D string attached) is not crucial in the penetration length for high energies, and
the result depends only on the energy $E$ and the virtuality $Q=\sqrt{E^2-k^2}$.

The result at the leading large $N_c$ and large coupling limit can be summarized by the following two lines\cite{Hatta:2008tx}:
\begin{itemize}
\item When $Q\gg Q_s\equiv (E T^2)^{1\over 3}$ (Diffusion regime), 
$\Delta x \sim {1\over T}\left(E\over Q\right)^{1\over 2}$.
\item When $Q\ll Q_s$ (Geodesic regime), $\Delta x\sim {1\over T}\left(E\over Q_s\right)^{1\over 2}\sim {1\over T}\left(E\over T\right)^{1\over 3}$.
\end{itemize}
If one fixes the virtuality $Q$ and increases the energy $E$, one eventually enters the geodesic regime ($Q\ll Q_s$), and the energy scaling of the penetration length becomes $\Delta x\sim E^{1\over 3}$\cite{Hatta:2008tx,Gubser:2008as,Chesler:2008uy}.
However, Ref.\cite{Arnold:2010ir} pointed out that the typical virtuality in real situations is given by $Q^2\sim E L^{-1}$ where $L$ is the size of the initial wave packet. 
If one fixes $L$ and keeps increasing $E$, one has $Q\gg Q_s$ and enters the diffusion regime, and the scaling  becomes $\Delta x\sim ({E\over Q})^{1\over 2}\sim E^{1\over 4}$.
We emphasize that the difference between the two scalings is a simple consequence of the different assumptions on $(E,Q)$, and does not mean any inconsistency.

A more challenging problem appears when we start to include subleading corrections in $1\over N_c^2$.
One such correction considered in Ref.\cite{Shuryak:2011ge} is the back reaction to the jet trajectory from the gravitational perturbations induced by its own past trajectory.
In electrodynamics, this corresponds to the Abraham-Lorentz-Dirac force, and there are similar computations for the gravitational back reaction as well.
The work done by the back reaction should match the gravitational radiation in 5D at large distance, and it is one way of understanding why it is subleading in $1\over N_c^2$.
The gravitational perturbation created by the jet trajectory is proportional to $\delta g\sim G_N\sim {1\over N_c^2}$. While the energy-momentum of the jet itself is given by holographic renormalization as $T^{\mu\nu}\sim {1\over G_N}\delta g\sim {\cal O}(1)$, the 5D gravitational radiation is proportional to $(\delta g)^2\sim {1\over N_c^4}$ which induces a metric perturbation of $\delta g\sim {1\over N_c^4}$ at large distance. This gives us a gauge theory energy-momentum flux of order $\delta T^{\mu\nu}\sim {1\over G_N}\delta g\sim {1\over N_c^2}$. 
In 5D, the massless retarded Green's function is non-zero inside the light-cone (in other words, the causally connected region), and the back reaction 
comes from the entire past history of the trajectory, not only from the instantaneous motion as it is the case in 4D. However, for the high energy limit $\gamma\gg 1$, the leading
contribution does localize at the instantaneous motion, and the detailed computation gives the energy loss ${dE\over dt}\sim {1\over N_c^2} E^2$\cite{Shuryak:2011ge}. 
This energy loss has a much stronger energy dependence than and will dominate over the leading $N_c$ results for any finite $N_c$.
For the realistic $N_c=3$, this brings a big question to the applicability of leading $N_c$ results in high energy jet quenching.
The point of this exercise is not to give a complete account of $1\over N_c^2$ corrections which have other sources than the back reaction we considered.
The purpose is to show that
the high energy limit and the large $N_c$ limit may not commute, which we also observe in the high $s$, small $t$ scattering before.

\section{Triangle anomaly}

To remind the audience of what the triangle anomaly is, consider a Dirac fermion field $\psi$. One can construct two currents out of it, the vector current $j_V^\mu=\bar\psi \gamma^\mu
\psi$ and the axial current $j_A^\mu=\bar\psi\gamma^\mu\gamma^5\psi$. While the vector current is the current of total fermion number which is strictly conserved,
the axial current is more like a current of spins as can be seen in a non-relativistic limit where $\vec j_A\sim \psi^\dagger\vec\sigma\psi$, and its conservation is much more fragile.
For example, a non-zero Dirac mass for $\psi$ breaks the conservation of axial current explicitly: $\partial_\mu j_A^\mu\propto m\bar\psi \gamma^5\psi$.
Even in the massless case where the axial current is conserved classically, there turns out to be a quantum mechanical violation that appears in and only in a 1-loop triangular 
Feynman diagram of $\langle j_A j_V j_V\rangle$ where $j_V$ vertices are coupled to the electromagnetism, which gives $\partial_\mu j_A^\mu={e^2N_c\over 16\pi^2}\epsilon^{\mu\nu\alpha\beta}F_{\mu\nu}F_{\alpha\beta}$\cite{Adler:1969gk,Bell:1969ts}.
The physics of triangle anomaly in low energy, chiral symmetry broken phase of QCD is captured completely by the Wess-Zumino-Witten Lagrangian.
The underlying axial-vector-vector three point function however has a richer form factor than just the low energy limit, and its possible physics consequences in high temperature and/or density regimes of QCD have not been fully explored.

One such phenomenon in a deconfined, chiral symmetry restored QGP is the Chiral Magnetic Effect (CME), which states that an axial (vector) chemical potential would induce
a vector (axial) current in the presence of a magnetic field: $\vec j_{V,A}={eN_c\over 2\pi^2}\mu_{A,V}\vec B$. 
In off-central heavy-ion collisions, the magnetic fields produced by the two heavily charged nuclei overlap constructively in the central collision region, and its magnitude can be
as large as $eB\sim m_\pi^2\sim 10^{19}$ Gauss, with a lifetime of $\lesssim 1$ fm, pointing perpendicular to the reaction plane. A possible experimental signature of the CME
that has been proposed is the following. In QGP, one expects QCD sphalerons to happen with a rate per volume given by $\Gamma\approx 30 \alpha_s^5 T^4$\cite{Moore:2010jd}. Each (anti-)sphaleron
increases (decreases) the axial charge by $2N_F$, and creates a local domain of non-zero axial chemical potential $\mu_A$. The CME then induces a vector current and hence a charge separation along the direction perpendicular to the reaction plane. A signal for this charge separation would be seen in the two point correlations of charged particles, $\langle \cos(\phi_1+\phi_2)\rangle$ where $\phi_{1,2}$ are the azimuthal angles from the reaction plane\cite{Voloshin:2004vk}. For the same charge correlation (i.e. $\langle\cdots\rangle_{++},\langle\cdots\rangle_{--}$), $\phi_1+\phi_2$ tends to be either $\pi$ or $-\pi$, and the $\langle \cos(\phi_1+\phi_2)\rangle$ should be negative. On the other hand, for the pairs of opposite charges, $\phi_1+\phi_2\approx 0$,
and $\langle \cos(\phi_1+\phi_2)\rangle$ should tend to be positive. Experiments at {\bf RHIC}\cite{Abelev:2009ac} and {\bf LHC}\cite{Selyuzhenkov:2011xq} confirmed this prediction, although there are other background effects unrelated to the magnetic field
that lead to the similar behavior\cite{Asakawa:2010bu,Bzdak:2009fc,Wang:2009kd,Pratt:2010zn}.  
A recent {\bf STAR} experiment of Uranium-Uranium collision sheds more light on this\cite{Wang:2012qs}. Uranium nucleus has a shape of ellipsoid, and the most central collision can still have a non-zero elliptic flow without creating a magnetic field, which is an ideal place to disentangle the physics originating from the elliptic flow alone and the physics of a magnetic field.
The $\langle \cos(\phi_1+\phi_2)\rangle$ correlations in the most central Uranium-Uranium collisions didn't show the behavior predicted by the CME, which supports that the previously  observed $\langle \cos(\phi_1+\phi_2)\rangle$ signals indeed originate from the magnetic field. 

\begin{figure}[t]
	\begin{center}
	\includegraphics[height=5cm]{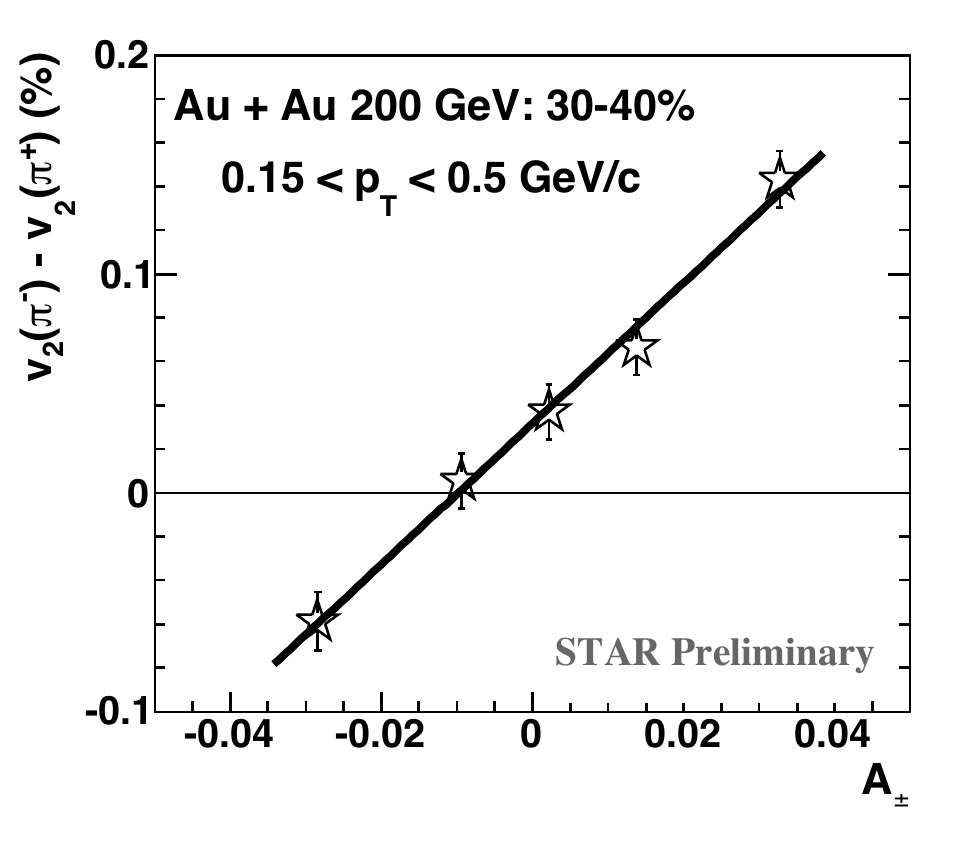}\includegraphics[height=5cm]{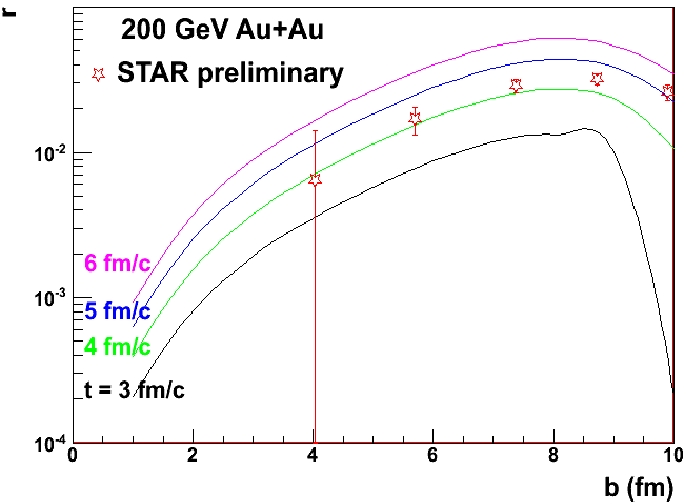}\end{center}
		\caption{ Result from {\bf STAR} \cite{Wang:2012qs} ({\it left}) confirming the linear dependency of $\Delta v_2$ on the charge imbalance $A_\pm\equiv{(N_+-N_-)\over(N_++N_-)}$ predicted by CMW.
		The slope $r\equiv{\Delta v_2\over A_\pm}$ as  we vary the impact parameter $b$ ({\it right}).
		Colored curves are the theory predictions based on CMW with different life-times of the magnetic field \cite{Burnier:2011bf,Burnier:2012ae}.}\label{fig1}
\end{figure}
Although the CME exists only in the presence of a chemical potential, there is a more general charge transport phenomenon arising from triangle anomaly: the Chiral Magnetic Wave (CMW).
It is a result of an interplay between the two versions of the CME, $\vec j_{V,A}={eN_c\over 2\pi^2}\mu_{A,V}\vec B$.
Suppose there was a local fluctuation of axial charge. Via the CME, there would be a local vector current, which brings up a local vector charge fluctuation in the vicinity. The other version of CME acting on this vector charge fluctuation would then induce another local axial charge fluctuation, and one easily recognizes that this interplay can continue by itself. 
The result is a sound-like propagation of chiral charge fluctuations with a dispersion relation $\omega=\pm v_\chi k-iD_L k^2+\cdots$, where the velocity $v_\chi={e^2 N_c \over 4\pi^2 \chi}$
is given in terms of the susceptibility $\chi$, and the sign (and hence the propagation direction with respect to the magnetic field) depends on the chirality of the fluctuations.
Recall that the original vector (axial) charge is given by $Q_{V,A}=\pm Q_L+Q_R$ in terms of the chiral charges $Q_{L,R}$. The CMW states that any profile of $Q_L$ moves in one direction along the magnetic field, while $Q_R$ moves in the opposite direction. We emphasize again that no background chemical potential is needed for the CMW.

A possible experimental signature of the CMW was proposed in Ref.\cite{Burnier:2011bf,Burnier:2012ae}. The QGP created in heavy-ion collisions has a small net vector charge chemical potential of $\sim 20$ MeV for $\sqrt{s}=200$ GeV. This vector charge can be thought of as a sum of equal amount of left- and right-handed charges via $Q_{V,A}=\pm Q_L+Q_R$ and $Q_A=0$ in average. 
The CMW acting on these charges move them toward the poles of the fireball away from the reaction plane, and recalling that the vector charge is a sum of the two chiral charges,
this gives an excess of vector charge in the poles, whereas there will be a depletion on the reaction plane, resulting a net electric quadrupole moment developed.
This would lead to a charge dependent elliptic flows of pions, $\Delta v_2=v_2(\pi^-)-v_2(\pi^+)>0$. Naturally, the electric quadrupole moment and the resulting $\Delta v_2$ linearly
depend on the initial charge asymmetry measured by $A_\pm={(N_+-N_-)\over (N_++N_-)}$. Recently, {\bf STAR}\cite{Wang:2012qs} performed an analysis of $\Delta v_2$ and confirmed the prediction by the CMW such as the linear dependency on $A_\pm$ as well as the dependency of the slope on the impact parameter. See Figure \ref{fig1}.

%\section*{References}

\end{document}